\documentclass[twocolumn,
 amssymb, amsmath,%
 aip,cha,%
]{revtex4}

\usepackage{amsmath}

\usepackage[dvips]{graphicx}

\usepackage{docs}%
\usepackage{bm}%

\begin{document}

\title{Continuous model for pathfinding system with self-recovery property} 
\author{Kei-Ichi Ueda}

 \affiliation{Graduate School of
Science and Engineering, University of Toyama, 
Toyama 930-8555, Japan}
\author{Yasumasa Nishiura}
 \affiliation{WPI Advanced Institute for Materials Research,
Tohoku University, Miyagi, 980-8577 Japan
}
\author{Yoko Yamaguchi}
 \affiliation{Neuroinformatics Japan Center, RIKEN Brain Science 
Institute, Saitama, 351-0198 Japan
}
\author{Keiichi Kitajo}
\affiliation{RIKEN, BSI-TOYOTA Collaboration Center; RIKEN Brain 
Science Institute, Saitama, 351-0198 Japan}

\begin{abstract}
This study propose a continuous pathfinding system based on coupled
 oscillator systems. We consider acyclic graphs whose vertices
 are connected by unidirectional edges. 
The proposed model autonomously finds a path connecting two
 specified vertices, and the path 
is represented by phase-synchronized oscillatory solutions. 
To develop a system capable of self-recovery, that is, a
 system with the ability to find a 
 path when one of the connections in the existing
 path is suddenly removed, we implemented three-state Boolean-like regulatory rules
 for interaction functions. We also demonstrate that appropriate installation of inhibitory interaction
 improves the finding  time. 
\end{abstract}

\maketitle
\section{Introduction}\label{sec:intro}
In biological systems, a large number of elements, such as cells and
organs,  interact with each other and adapt system-wide behavior spontaneously
in response to changes of environment and of physical constraints. 
Such spontaneous and adaptive dynamics have been studied in the framework of 
collective dynamics \cite{watt}. 
To clarify the mechanism of adaptability, however,
we need to investigate how the dynamics of interactions between elements should depend on the global behavior to 
establish adaptability. 

Collective dynamics has  
been studied as a distributed system to investigate how 
global structures emerge from the 
interaction of elements.
Many types of distributed systems have been modeled by using
differential 
equations, and the system behaviors have been described by appropriate
basin switching between attractors \cite{free}. 
For instance, previous studies investigated retrieval of a specific pattern by neural oscillator networks \cite{aoya,hopf}.
 In these studies, the stable pattern was predicted by 
Lyapunov function analysis, and the basin 
switching was accomplished by controlling the potential profiles. 
Complex dynamics observed in brain systems have been studied in terms of
chaotic itinerancy.  
Transient behavior between ordered and disordered states has been shown in some coupled oscillator systems. \cite{kane,kane2}. 

The aim of this study is to propose a
distributed system showing appropriate basin switching by using the 
pathfinding problem.  
Pathfinding strategies have been proposed in both theoretical 
and experimental contexts. 
In laboratory experiments, pathfinding algorithms based on 
self-organization processes have been studied
\cite{stei,dori,naka}. 
In \cite{tero}, a continuous model for the experimental results of 
\cite{naka} has been proposed.   
In their model, a conservation mass has been assumed, which 
allows the system to find the shortest path \cite{miya}. 

In the present study, we modify the loop searching system proposed in 
\cite{ueda} to apply it to the pathfinding problem in the framework of a
distributed system;  that is, no conservation mass is assumed. 
The path connecting the start and goal points is represented by a 
loop path containing both the start point and the goal point. 
In such a network structure, a pathfinding problem can be regarded as a loop
finding problem. 
Therefore, we propose a two-layer network system to construct such loop
networks.  
In addition, in
order to  
improve the finding time of the desired path, the effects of interactions between 
the dynamics of the two layers are considered.
The presented system shows the following three properties.
(i) The system spontaneously finds a path connecting the start and goal nodes 
if such a path exists. (ii)
The system shows the self-recovery property, that is, the system finds
another possible path when the existing path is broken 
due to the removal of paths.  (iii) In a tree structure network, 
the finding time increases in proportion to the depth of the
tree.

\section{Model}
The graphs we consider in this study have unidirectional edges
between vertices and are acyclic. 
The graph has two special vertices referred to as the start vertex 
and the goal vertex. 
In this section, we propose a construction procedure for a system that 
shows properties (i) to (iii). 
The path is defined by a stable
solution of the system.
To avoid confusion, we use the terms,  {\it vertex} and {\it edge} for the 
given graph and {\it node} and {\it link} for the same properties of the system.

\begin{figure*}[htbp]
 \begin{center}
  \includegraphics[keepaspectratio=true, width=0.70\textwidth]{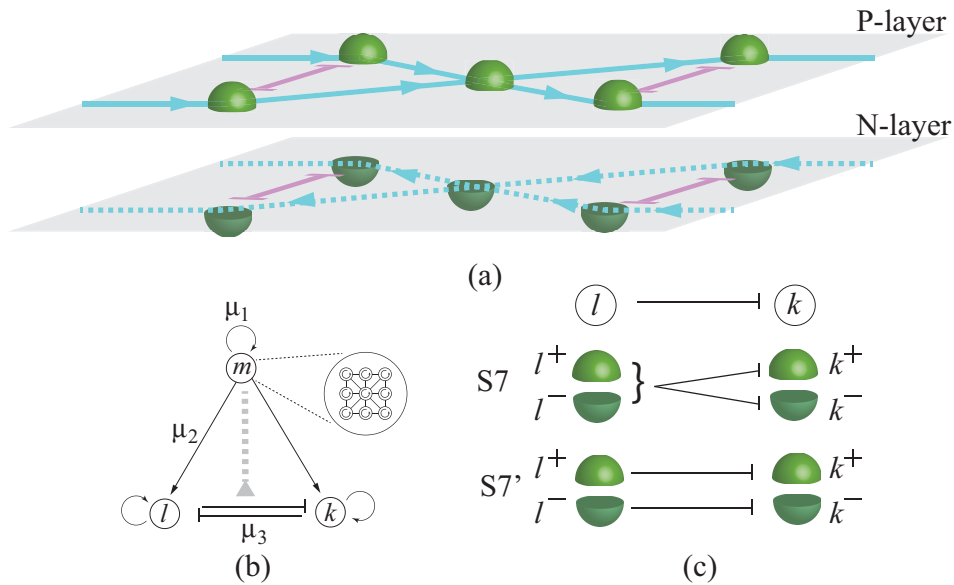}
\end{center}
 \caption{
(a) Schematic figure of the network structure of the presented
 system. Each node consists of a group of oscillators. 
From (S2), the directional interaction in the N-layer is opposite those in
 the P-layer.
(b) Bidirectional inhibitory links exist at the branching
 point from (S2). The activation of the inhibitory
 interaction between nodes $k^\pm$ and $l^\pm$ is regulated by the state
 of node $m^\pm$.   
(c) The inhibitory interaction for the current (S7) and previous (S7')
 system. 
}
\label{fig:bilayer}
\end{figure*}

The assumptions for the system are given as follows. 
\begin{itemize}
\item[(S1)] 
We assume two layers exist, and a pair of nodes is placed on 
the layers at every vertex position (Fig.\ref{fig:bilayer}).  
For notational convenience, 
we refer to  one of the layers as the P-layer (positive) and to the other one as the 
N-layer (negative). The node numbers corresponding to the  $k$th vertex in the
	    P-layer and in the N-layer are given by $k^+$ and $k^-$, respectively.

\item[(S2)] 
The excitatory links in the P-layer and the N-layer take place along the forward and
	    backward direction of the edges, respectively.  That is, 
if the network of the P-layer has an excitatory link from node $l^+$ to
 	    node $k^+$, then the network of the N-layer has an excitatory link from
	    node $k^-$ to node $l^-$.
Bidirectional inhibitory links take place at every branching point of
	    excitatory links. Additional assumptions for the inhibitory interactions
	    are given in (S7) and (S7'). 
\item[(S4)]  Each node contains a large number of oscillators, and the state of
each node is determined by the dynamics  of the group oscillators corresponding to the node. 
\item[(S5)] Each P-layer and N-layer has two special nodes,
	    corresponding to 
	    the start vertex and the goal vertex, referred to as the
	    start node  and the goal node, respectively. The start nodes of the P-layer and the N-layer
	    are labeled $k_s^+$ and $k_s^-$, respectively. Similarly, 
	    the goal nodes of the P-layer and the N-layer
	   are labeled $k_g^+$ and $k_g^-$, respectively.
	    The network has an excitatory link from 
	    $k_s^-$ to $k_s^+$ and from $k_g^+$ to $k_g^-$. 
\item[(S6)] 
The network in the P-layer and the N-layer without the two links placed at the start
	    and goal nodes is acyclic. 
\end{itemize}

If a connecting path from the start vertex to the goal vertex exists, then a connecting
path from the start node to the goal  
node exists in the P-layer and a connecting path from the goal node to the start node exists in
the N-layer from (S2). Therefore, from (S5),  loops of excitatory links are 
formed; these loops contain nodes $k_s^\pm$ and $k_g^\pm$. 

\subsection{Equations}
The governing equations are as follows:
\begin{equation}\label{eqn:model}
\begin{aligned}
 &\dot{u}^j_{k^\pm} = [u^j_{k^\pm}(1-u^j_{k^\pm})(u^j_{k^\pm}-p) -v^j_{k^\pm}]/\tau_j \\
 &\hspace{0cm}+\mu_1
 F(s_{k^\pm};\bar{s})+\mu_2\sum_{l^\pm\in \Lambda^\pm} 
a_{l^\pm,k^\pm}F(s_{l^\pm};\bar{s}) + w(t),\\
 &\dot{v}^j_{k^\pm} = \varepsilon(u^j_{k^\pm}- q v^j_{k^\pm} +
 r)/\tau_j\\
&\hspace{0cm}
 +\mu_3\sum_{m^{\pm}=1^\pm}^{K^\pm}\sum_{l^\pm\in\Lambda^\pm}a_{m^\pm,l^\pm}a_{m^\pm,
 k^\pm} \\ 
&\hspace{2cm}  \times F(s_{m^{\pm},k^{\pm},l^\pm};\bar{s})F(s_{l^\pm}+\sigma s_{l^\mp};\gamma \bar{s})\\
 &s_{k^\pm}=\frac{1}{J}\sum_{j=1}^{J}u^{j}_{k^\pm},\quad
k^\pm=1^\pm,2^\pm,\dots,K^\pm,\ j=1,2,\dots,J,
\end{aligned}
\end{equation}
where $\Lambda^\pm:=\{k\ |\ 1^\pm\leq k \leq K^\pm, k = k^\mp\}$,  $t$ is dimensionless time, a dot above a variable indicates the derivative of that variable with
respect to $t$, and $\mu_i$ ($i=1,2,3$) and $\gamma$ are positive constants. 
The dynamics of each oscillator are described by the FitzHugh-Nagumo
equation. $K^+$ (or $K^-$) is the total number of vertices. 
From (S4), each node consists of $J$ oscillators and 
the variables $u_k^j$ and $v_k^j$ correspond to the activator and inhibitor, respectively,
of the $j$th oscillator in node $k$. 
The state of each node $k^\pm$ is determined by $s_{k^\pm}$, the average
of $u^j_{k^\pm}$ ($j=1,2,\dots, J$) in node $k^\pm$. 
The strength of the self-feedback  $\mu_1$ controls the degree of 
phase synchronization of the node, where large and small values of
$\mu_1$ induce 
synchronization or incoherent oscillation, respectively. 
Parameters $\mu_2$
 and $\mu_3$ correspond to the strengths of excitatory and inhibitory
connections, respectively [Fig.\ref{fig:bilayer}(b)].

We assume that the interaction function $F$ has a threshold 
for activation; thus, the state of the connection switches dynamically 
between the on and off states depending on the state of the node. 
The regulation of the on-off switching
of the connecting nodes depends on $s_k$ and is defined by the
Heaviside function with a
threshold $\bar{s}$, where $F(s_k;\bar{s})=1$ for $s_k>\bar{s}$ and
$F(s_k;\bar{s})=0$ for $s_k\leq \bar{s}$. 
The interactions affect 
all elements uniformly; that is, they are independent of $j$. 
The implementation of the Heaviside function is needed to realize condition C3, 
given later.

Excitatory interaction directed from node $l$ to node
$k$ is expressed as $a_{l,k}$, where 
$a_{l,k} =1$ and $a_{l,k} =0$ indicate the presence
and absence of such interactions, respectively.  
Since the start and goal nodes have a connecting
path  from the N-layer to the P-layer 
and a path from the P-layer to the N-layer from (S5), 
$a_{k_s^-,k_s^+} = a_{k_g^+,k_g^-}=1$. 
It is noted that $a_{k^\pm,k^\mp}=0$ except for $k^\pm\not=k_{s}^\pm$ 
and $k^\pm\not=k_{g}^\pm$ and that $a_{k^\pm,k^\pm}=0$ for all $k^\pm$. 

From (S2), the system has inhibitory interactions at 
every branching point of the network (see condition C2). 
The activation of the inhibitory interaction between nodes $k$ and $l$ 
is also regulated by $s_m$ [Fig.\ref{fig:bilayer}(b)]. 
We use the notation 
$s_{m,k,l}$ to indicate that  node $m$ regulates activation of inhibitory interactions between nodes $k$ and $l$.

Assumptions (S7) and (S7') correspond to the inhibitory
functions of the presented system and the previous system, respectively. 
\begin{itemize}
 \item[(S7)] For the presented system, we take $\sigma=1$.  
\end{itemize}
Except for the links at the start and goal nodes, 
the dynamics of the P-layer and the N-layer are coupled through this inhibitory
interaction. 
From (S7), the inhibitory interactions are activated when 
$s_{l^\pm}+s_{l^\mp}$ is larger than $\gamma \bar{s}$ and
$s_{m^\pm_,k^\pm,l^\pm}$ is larger than $\bar{s}$. 

To compare the finding time of the presented system with that of the previous
system, we use the previous system in Section \ref{sec:results}. Since 
the previous system is fundamentally the
same as (\ref{eqn:model}) 
with $\sigma=0$ and $\gamma=1$, 
we assume (S7') when we examine numerical experiments of the previous system.
\begin{itemize}
 \item[(S7')] For the previous system, we take $\sigma=0$ and
	      $\gamma=1$. 
\end{itemize}
We assume (S7) unless otherwise mentioned. 

 For simplicity, we assume that the parameters $p$,
$q$, $r$, and $\varepsilon$ 
are independent of $k^\pm$ and we set $p=0.02$, $q=1.0$, $r=-0.04$,
and $\varepsilon=0.01$; $w$ is a small amount of random noise in the
interval  $[0,0.05]$; the time constants $\tau_j$ take random values from
the interval between  
$\tau_{\min}$ and $\tau_{\max}$, where the values of $\tau_{\min}$ and $\tau_{\max}$
 are set to $(\tau_{\min},\tau_{\max})=(6.0, 6.5)$. The distribution of $\tau_j$
 is the same for all nodes; that is, it is independent of $k^\pm$.

\subsection{Regulatory rules}
The regulatory rules proposed in \cite{ueda} are employed for the presented
model. Each node can have three 
qualitatively different states: 
synchronized oscillation with a large amplitude, synchronized oscillation 
with a small amplitude, and incoherent oscillation. 
We refer to these states as large-amplitude synchronized oscillation (LSO), 
small-amplitude synchronized oscillation (SSO) and incoherent
oscillation (INC), respectively. 
The state LSO means that $s_k$ exceeds $\bar{s}$ for every
oscillation. 
The use of the FitzHugh-Nagumo equation admits the presence of SSO  when 
inhibitory interactions are exerted. 
By regarding LSO as the on state and both SSO and INC as the off state, 
the proposed system can be regarded as a Boolean network.

\begin{figure}[htbp]
 \begin{center}
  \includegraphics[keepaspectratio=true, width=0.50\textwidth]{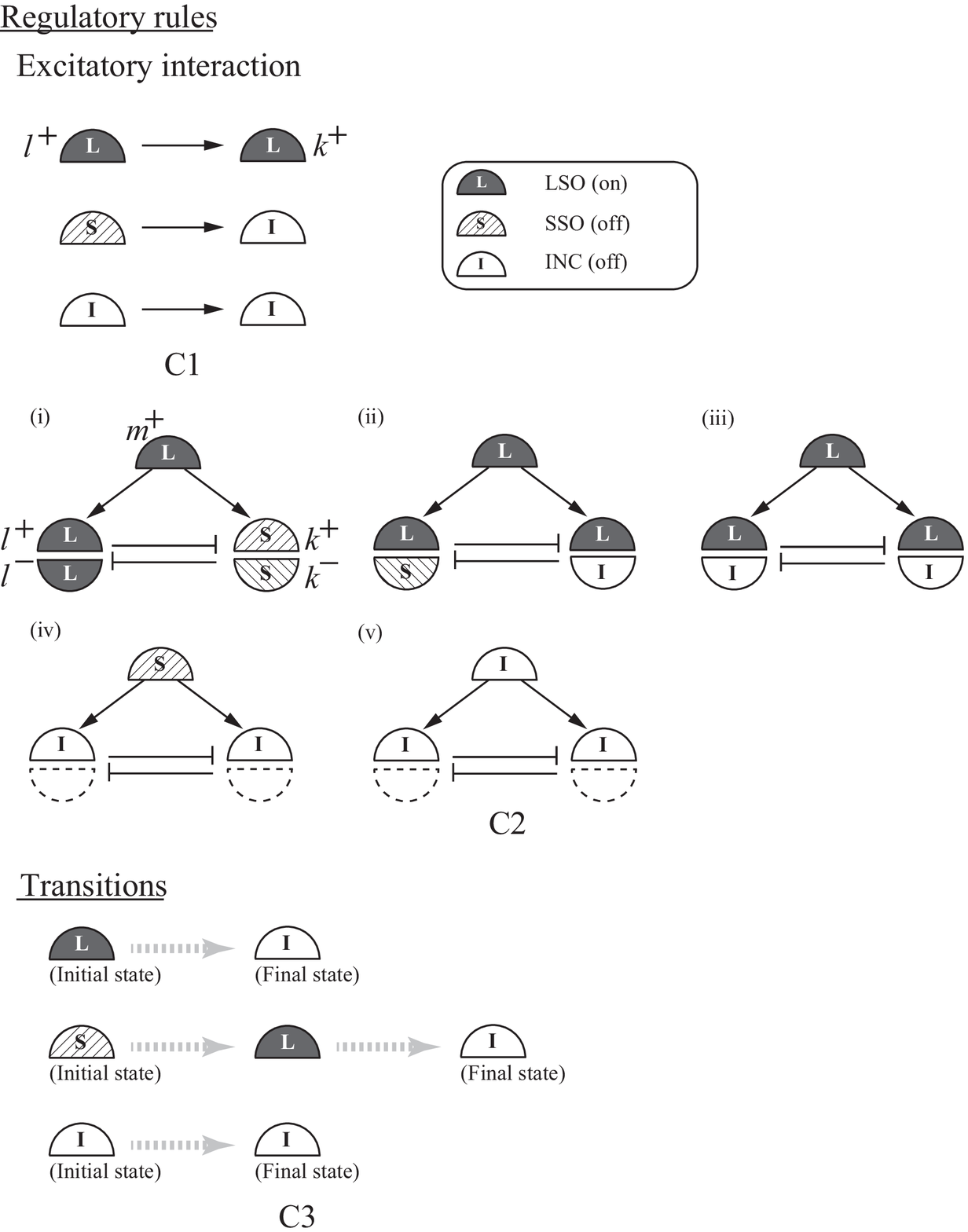}
\end{center}
 \caption{(C1) Regulatory rules for excitatory and inhibitory
 interactions.  
(C2) Possible patterns of nodes in the P-layer for the system with (S7) at
 the branching point of  excitatory links. (i) The 
 nodes $k^+$ and $k^-$ are inhibited when nodes $l^+$, $l^-$,
 and $m^+$ are in an LSO state. (ii)(iii) The state of node $k^+$ and 
 $l^+$ becomes LSO when the state of node $m^+$ is the on state and
 nodes $k^-$ and $l^-$ are in
 off states (SSO or INC). (iv)(v) The states of nodes $k^+$ and
 $l^+$ are INC when node $m^+$ is in the off state,
 independent of the state of nodes in the N-layer. For all patterns, 
the roles of $k^+$ and  $l^+$ can be interchanged. The possible patterns for
 the N-layer are given similarly.
(C3) Transient behavior when a node receives no
 input. The solution converges to an
INC for any initial condition. 
 Node starting with an SSO approaches a quasi-stable LSO during a
 finite time interval and finally converges to INC.}
\label{fig:condition}
\end{figure}

The parameter values $\mu_i$ are set so that the regulatory rules and
the  conditions (C1,C2, and C3 in Fig. \ref{fig:condition}) are
satisfied. The parameter values are set to   
$(\mu_1,\mu_2,\mu_3)=(1.8\times 10^{-3}, 0.06, 0.07)$ and
$\bar{s}=0.825$. As discussed later, the state of a node shows INC when 
no interactions from other nodes are exerted. 
The value of $\gamma$ is taken from  an appropriate regime satisfying
the conditions. 
Here we discuss the regulatory rules for nodes in the P-layer. 
The same argument can be applied, {\it mutatis mutandis}, to the regulatory rules for the nodes in the N-layer. 

Since the interactions exert uniformly on the oscillators
belonging to the same node, phase resetting of oscillators is observed
due to the excitatory interactions.
For the given parameter values, the LSO state is propagated along  
the nodes connected by excitatory links. As shown in
Fig.\ref{fig:condition}, the state of node $k^+$ 
 becomes LSO when that of node $l^+$ is LSO 
[Fig.\ref{fig:condition} (C1)]. 
The node $k^+$ becomes INC when  
the node $l^+$ is in one of the off states (SSO or INC) since 
no interactions are exerted on node $k^+$.

At the branching point of excitatory links, we can see the 
competition between the excitatory and inhibitory interactions. 
For the inhibitory interactions, the regulatory rules are determined 
by the states of the node in both layers. 
The inhibitory interactions from node $k^{\pm}$ to node $l^{\pm}$ become
active when both node $k^+$ and node $k^-$ are in the on state, or 
$s_{k^+}+s_{k^-}> \gamma \bar{s}$. Thus, 
the inhibitory interactions from node $k^{\pm}$ to node $l^{\pm}$ 
become inactive when either the state of 
node $k^+$ or that of node $k^-$ is one of the off states (SSO or INC). 

When the state of node $m^+$ is LSO, the  
excitatory interactions are exerted on node $k^+$ and node $l^+$ from 
nodes $m^+$. In the case that the states of node $l^-$ and $l^+$ are LSO
as shown in Fig.\ref{fig:condition} (C2)(i),  
the states of node $k^+$ and $k^-$ 
become SSO due to the inhibitory interactions. 
The role of $k^+$ and  $l^+$ is interchanged when the states of node
$k^+$ and $k^-$ are LSO. 
In the case that node $l^-$ and $k^-$ are both in one of the off states,  
as shown in
Fig.\ref{fig:condition} (C2)(ii) and (iii), the state of nodes
$k^+$ and $l^+$ becomes LSO. 
When the state of node $m^+$ is one of the off states, 
since $F(s_{m^+,k^+, l^+})=0$ and the inhibitory interactions are
inactive, 
the nodes $k^+$ and $l^+$ become 
INC, independent of the state of the nodes in the N-layer [Fig.\ref{fig:condition} (C2) (iv) and (v)]. 

\begin{figure}[htbp]
 \begin{center}
  \includegraphics[keepaspectratio=true, width=0.40\textwidth]{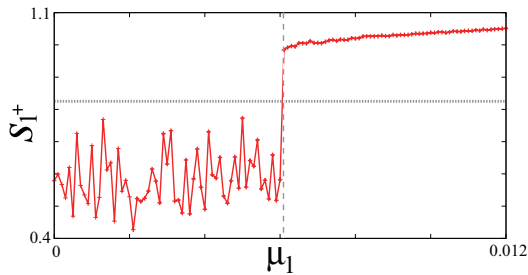}
\end{center}
 \caption{Bifurcation diagram of an isolated node 
 ($\mu_2=\mu_3=0$). The vertical axis indicates the maximum value of
 $s_{1^+}$. 
The horizontal dotted line indicates the threshold $\bar{s}$. The limit point is
 $\mu_1\approx 0.006$ (vertical dashed line).}
\label{fig:bif}
\end{figure}

Parameter $\mu_1$ is set so that the single node ($\mu_2= \mu_3=0$) 
shows transient  behavior SSO $\to$ LSO $\to$ INC
when  SSO is taken as an initial data point (C3).
Figure \ref{fig:bif} shows the bifurcation diagram of the 
single node (node $1^+$). 
It is observed that such a state transition from LSO to INC occurs at a limit
point, $\mu_1\approx 0.006$.  
By moving $\mu_1$ 
off the limit point of LSO, transient behavior from LSO to INC is observed, and the transition time increases as the distance between
$\mu_1$  and the limit point becomes smaller. 
Note that the transient time of LSO becomes longer as the
distance from the limiting point becomes smaller. 
The threshold $\bar{s}$ is taken to be sufficiently
large to  ensure that the values of $s_{k^\pm}$ are smaller
than $\bar{s}$, and the state of the node converges to INC 
when no input is received (see the horizontal dotted 
line in Fig.\ref{fig:bif}).

In Section \ref{sec:results}, we can observe that a loop chain of LSO states can
be stable  due to the regulatory rule (C1). The self-recovery process is
established owing to the 
regulatory rule (C2) with the transient behavior (C3).

\begin{figure*}[htbp]
 \begin{center}
  \includegraphics[keepaspectratio=true, width=0.70\textwidth]{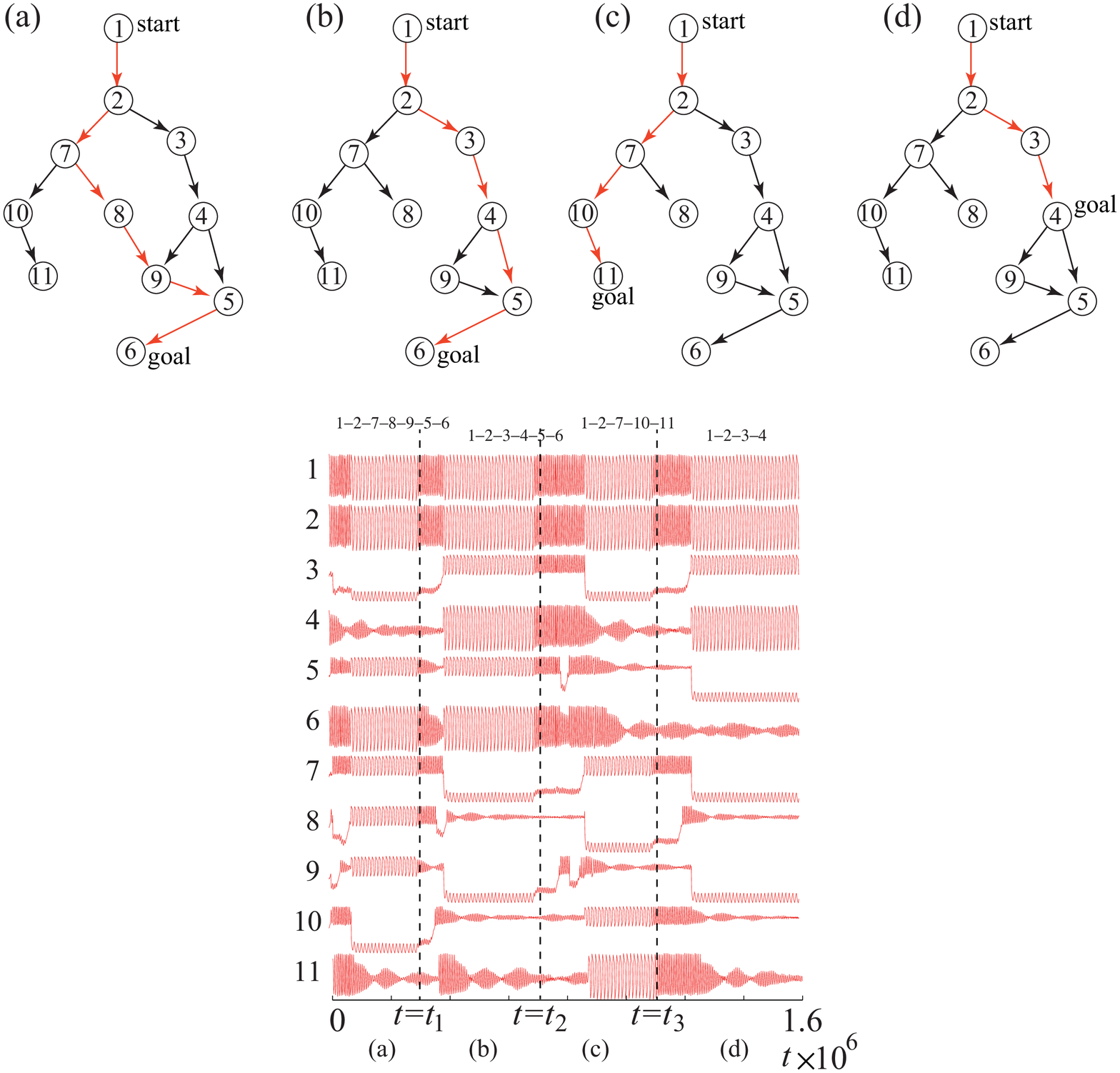}
\end{center}
 \caption{ Upper: The graphs used in numerical experiments.  
Bottom: Time sequences of $s_k^+$. 
The sequences of numbers at the top of the figures are the selected vertices. 
For example, for $t<t_1$, vertices 1, 2, 5, 6 7, 8, and 
9 are selected.}
\label{fig:perturb}
\end{figure*}

\section{Results}\label{sec:results}
In this section, we confirm that the presented system shows the
properties (i)-(iii) listed in Section \ref{sec:intro}.
A path is defined as a chain of LSO states belonging to
the corresponding nodes in both the P-layer and the N-layer. 
In addition, we demonstrate that the presented system tends to choose shorter paths
when multiple paths exist.

\subsection{Pathfinding}
We test the presented system in the network shown in 
Fig.\ref{fig:perturb}. From (S2), bidirectional inhibitory
links are placed between nodes
$3^\pm-7^\pm$, $8^\pm-10^\pm$, $5^\pm-9^\pm$, 
and  $4^\pm-8^\pm$.  
Initially, we take the start and goal vertices as vertices 1
and 6, respectively; that is, $k_s^\pm=1^\pm$ and $k_g^\pm=6^\pm$
[Fig.\ref{fig:perturb}(a)].  
At $t=0$, the network has three possible paths connecting the
start and goal vertices: path $1\to 2\to 3\to 4\to 5\to 6$,  path
$1\to 2\to 3\to 4\to 9\to 5\to 6$, and 
path $1\to 2\to 7\to 8\to 9\to 5\to 6$. 
In numerical results shown in
[Fig.\ref{fig:perturb}(a)], path $1\to 2\to 7\to 8\to 9\to 5\to 6$ is 
selected from the possible paths. 

To confirm the self-recovery 
property, edge 8$\to$9 of the existing edge is removed at $t=t_1$
[Fig.\ref{fig:perturb}(b)]; the parameter values of $a_{8^+,9^+}$ and 
$a_{9^-,8^-}$ are changed from 
$1$ to $0$ at $t=t_1$. 
Then, it is observed that the system can find a new path 
$1\to 2 \to 3\to 4\to 5\to 6$.  

The system can find a new path when the positions of the start and goal vertices are
changed. For example, in Fig.\ref{fig:perturb}(c)(d), it is observed
that the system finds a path
when the position of the goal vertex is changed from 6 to 11 at $t=t_2$
($a_{6^+,6^-}=0, a_{11^+,11^-}=1$)
and  from $11$ to $4$ at $t=t_3$ 
($a_{11^+,11^-}=0, a_{4^+,4^-}=1$). 
Due to the addition of the excitatory links at $t=t_3$, 
inhibitory links are added between $9^\pm-4^\pm$ and 
$5^\pm-4^\pm$ from (S2) and (S7). 
When the network has no connecting paths,  all of the node
states become INC
from the regulatory rules.

\subsection{Increasing the rate of finding }\label{sec:time}
\begin{figure}[htbp]
 \begin{center}
  \includegraphics[keepaspectratio=true, width=0.5\textwidth]{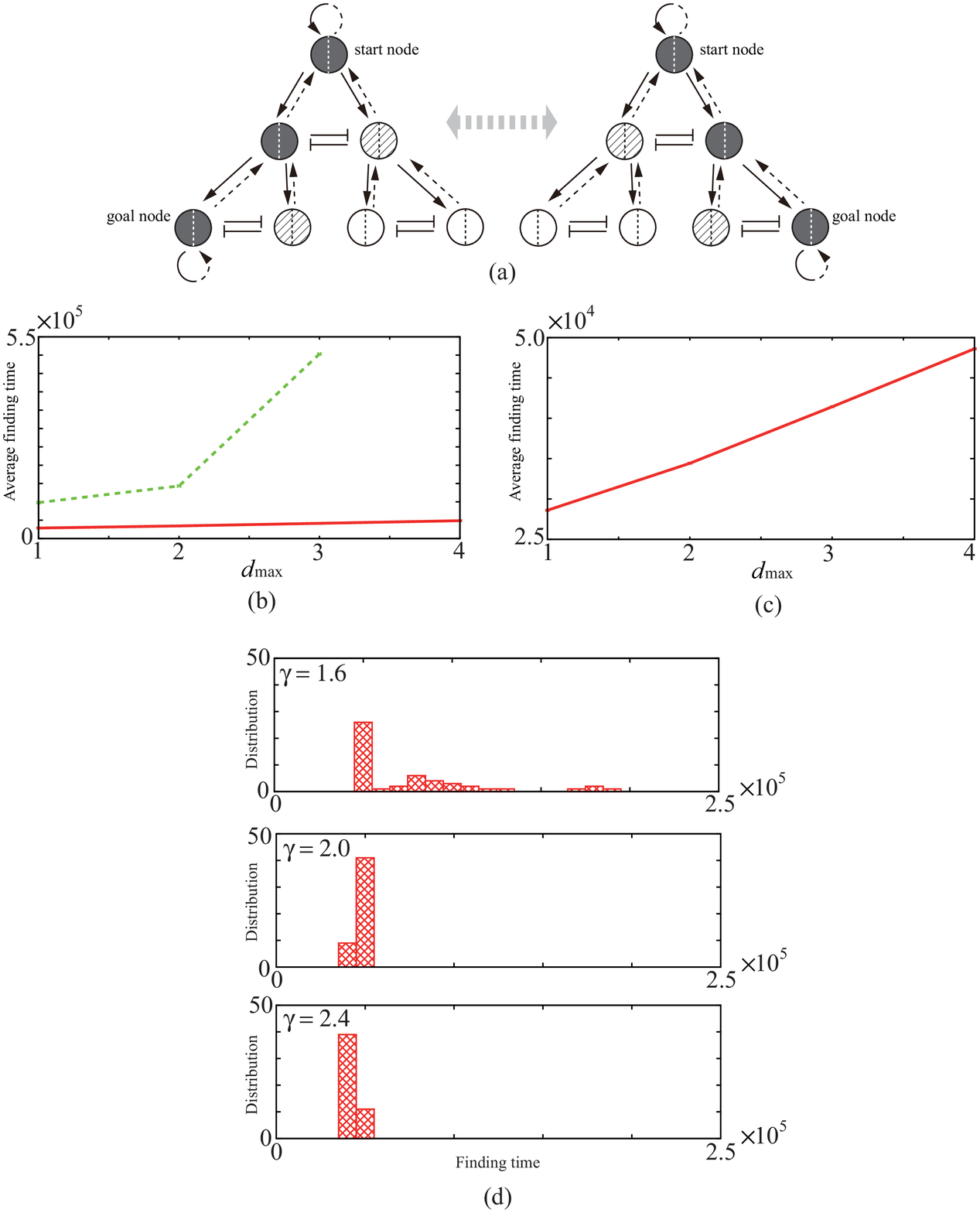}
\end{center}
 \caption{(a) Tree network of depth 2. The left and right halves of the
 circle indicate nodes in the P-layer and the N-layer, respectively.
The solid and dashed lines
 indicate excitatory links corresponding to the P-layer and the N-layer,
 respectively. 
The position of the goal alternates between the two types
 of networks, changing when the system has succeeded in finding a path. 
(b) The average finding time of the presented system (solid
 lines) with $\gamma=2$ and of the previous system (dashed line). 
(c) Magnified figure of (b). (d) The distribution of finding time of
 the presented system with $d_{\max}=3$ and $\gamma=1.6, 2.0, 2.4$. }
\label{fig:tree}
\end{figure}

In order to estimate the increase in the rate of 
finding as the node number increases, we use 
a hierarchical tree network, shown in Fig.\ref{fig:tree}:
$a_{k+2^{d-1},2k+2^d}=a_{k+2^{d-1},2k+1+2^d}=1$ 
($k=0,1,\dotsc,2^{d-1}-1$, $d=1,2,\dots,d_{\max}$), where 
$d_{\max}$ is the depth of the hierarchy. 
The start position is fixed at vertex 1,
and the goal vertex is changed between the leftmost vertex (vertex
$2^{d_{\max}-1}$) and the rightmost vertex (vertex $2^{d_{\max}}-1$)  
at the bottom of the tree when
the system has succeeded in finding a desired path. 
For example, in the case of $d_{\max}=2$ (Fig.\ref{fig:tree}), the path
contains vertices 1, 2, and 4 when 
the goal vertex is 4 and vertices 1, 3, and 7 when the goal vertex is 7. 
The average finding time is calculated across 50 trials.  

Now we compare the finding time for the presented system [(S7)] with 
$\gamma=2$ and the
previous system  [(S7')]  and $d_{\max}$ are taken as control parameters. 
In the case of the previous system, the finding time increases
substantially as $d_{\max}$ increases [Fig.\ref{fig:tree} (b)]. 
In contrast, the finding time increases at an almost constant rate for the 
presented system. 

Next, we investigate the dependence of $\gamma$ on the finding time. 
It is observed that the finding time is improved  and the distributions
tend to be narrow when $\gamma$ is
 increased for $\gamma \leq 2.4$ and $d_{\max}=1,\dotsc,4$. 
 Figure \ref{fig:tree} (d) shows the distributions of the finding time
 when $\gamma$ is taken as a control parameter with $d_{\max}=3$. 
We note that the distributions tend to be wide again for 
$\gamma\geq 2.6$. This implies that the large $\gamma$
basically decreases the finding time; however, a large $\gamma$ also decreases the reliability
when the vertex number increases.

\begin{figure*}[htbp]
 \begin{center}
  \includegraphics[keepaspectratio=true, width=0.85\textwidth]{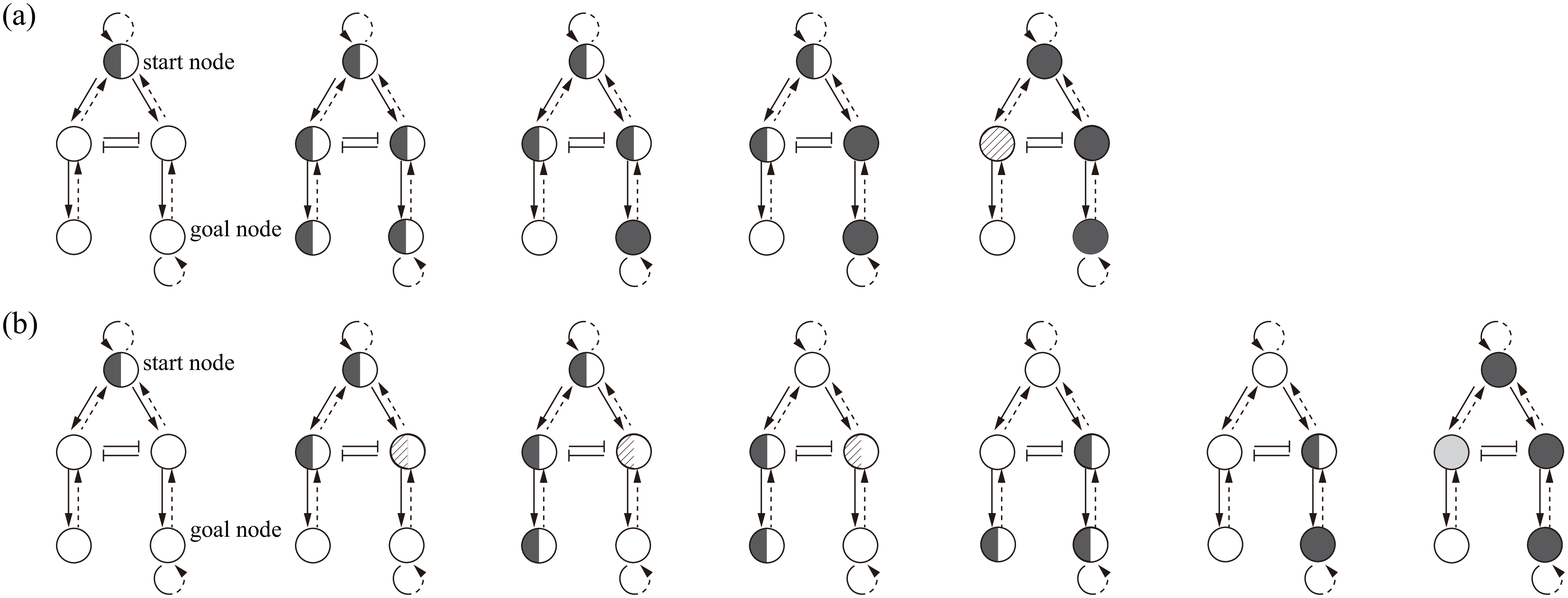}
\end{center}
 \caption{The finding processes of the presented [(a)] and 
previous [(b)] systems. 
(a) An LSO wave initially propagates to the wrong 
 path, inhibitory interaction occurs in the second panel from the left, and finally
 the system converges to the desired solution.
(b) Inhibitory interaction occurs after an LSO wave goes through an endpoint
 (4th panel). The left and right halves of the circle indicate the nodes
 in the P-layer and the N-layer, respectively. 
}
\label{fig:hikaku}
\end{figure*}

The difference of the finding time between the two systems comes from the timing of the activation of the inhibitory 
interaction. The inhibitory interactions play a role
in path selection at the branching point of excitatory links. 
In the presented system, the activation of the inhibitory interaction at
node $k^\pm$ is regulated by both $s_{k^+}$ and $s_{k^-}$. 
This means that the inhibitory interaction for a large $\gamma$ becomes active 
(i.e., path selection initiates) after an LSO 
wave propagates to, at least,  either the start node or the goal node. 
In the example network in Fig.\ref{fig:hikaku} (a),  
the path selection initiates  
after the system succeeds in finding a desired path. 
This prevents the system from selecting wrong paths, and thus results in a decrease in finding
time. 
In contrast, for the previous 
system, since the 
inhibitory interactions activate when either $s_{k^+}$  or
$s_{k^-}$ is larger than $\bar{s}$, 
the selection of a path at branching points occurs before the
system has found both the start and goal vertices, as shown in the example network 
in Fig.\ref{fig:hikaku} (b). 

\subsection{Ability of shortest path selection}
As discussed in Section \ref{sec:time}, for the presented system,
the inhibitory interactions become active when an LSO wave, 
which has already passed through either the start node or the goal node,  
first reaches the branching point. 
Therefore, it is expected that the system chooses a path for which the
distance between the branching point 
and the start or goal node is the shortest; this is, 
the node distance between the branching point and the start (or goal) node
is smallest. 

\begin{figure}[htbp]
 \begin{center}
  \includegraphics[keepaspectratio=true, width=0.40\textwidth]{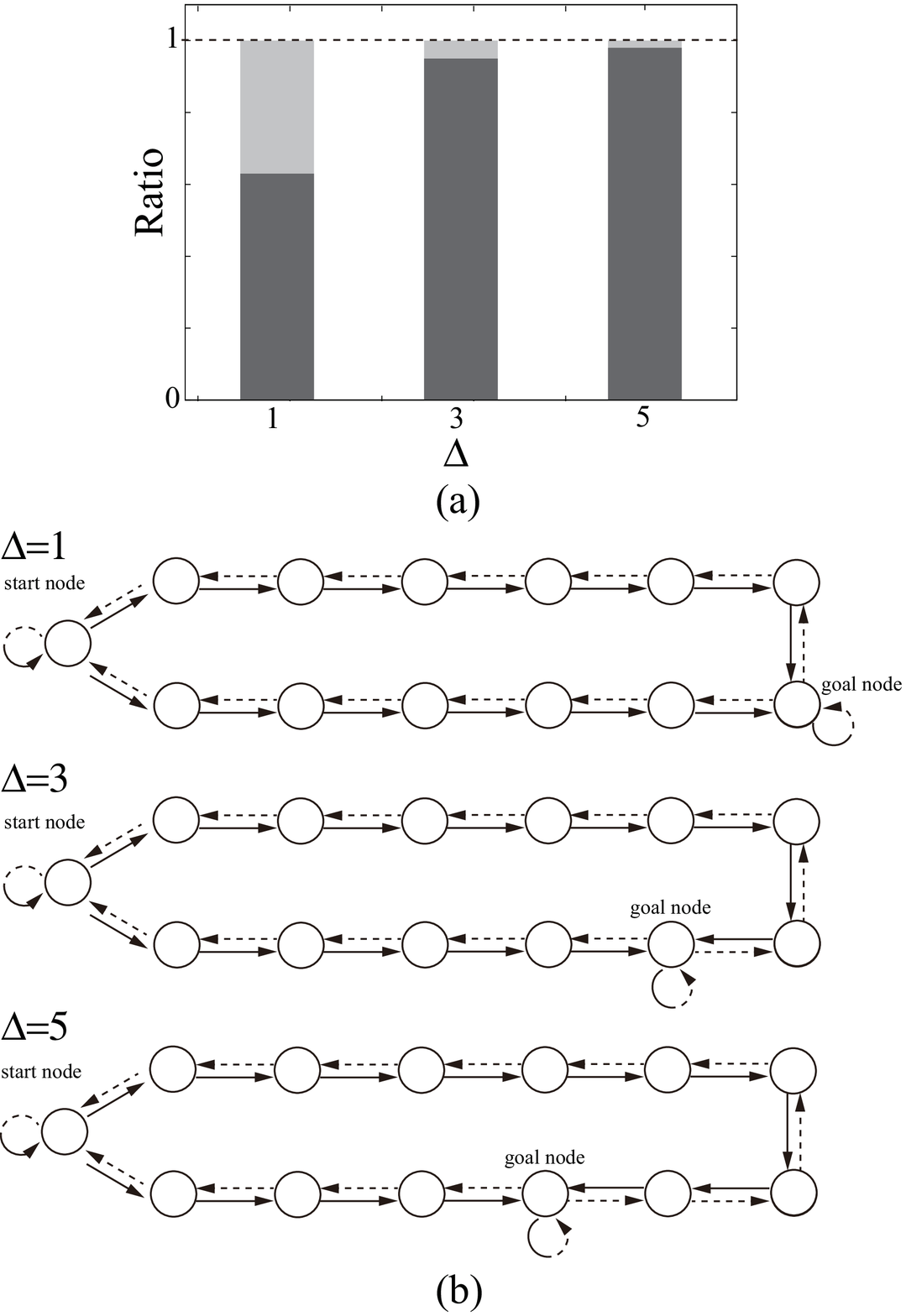}
\end{center}
 \caption{(a) The ratio of success in finding short paths for the
 network in (b). 
Black and gray bars indicate the finding ratio of shorter
 and longer paths for $\Delta =1,3,5$, respectively. 
(b) Network structures. We fix $K^\pm=13$. 
 }
\label{fig:shortest}
\end{figure}

The ratio of the chosen path to the shortest path is calculated for a 
network shown in Fig.\ref{fig:shortest}, where the distance of the path is
defined by the number of nodes in the path. 
The difference of the node number $\Delta$ is taken as a
control parameter. 
As initial data, the state of the start node is set to LSO and 
the other nodes are set to INC. The number of trials is 
100 with different random seeds. 
It is confirmed that the presented system with $\gamma=2$ and $K^\pm=13$
selects shorter paths more
frequently as $\Delta$ increases. 

\section{Discussion}
In this study, by building on a previous system \cite{ueda}, we have
presented a pathfinding system. The finding process uses only local 
interaction between nodes; that is, the presented system is in a
class of distributed systems. The system inherits the self-recovery
property from the previous system. 
In fact, we have demonstrated that the system can find a new
path when perturbation of the existing paths occurs.
We further modify the inhibitory interactions to improve the 
finding time of the desired path. It has been  numerically confirmed 
that the finding time increases linearly as the depth of the hierarchy 
of the tree-like network increases. 

\begin{figure}[htbp]
 \begin{center}
  \includegraphics[keepaspectratio=true, width=0.5\textwidth]{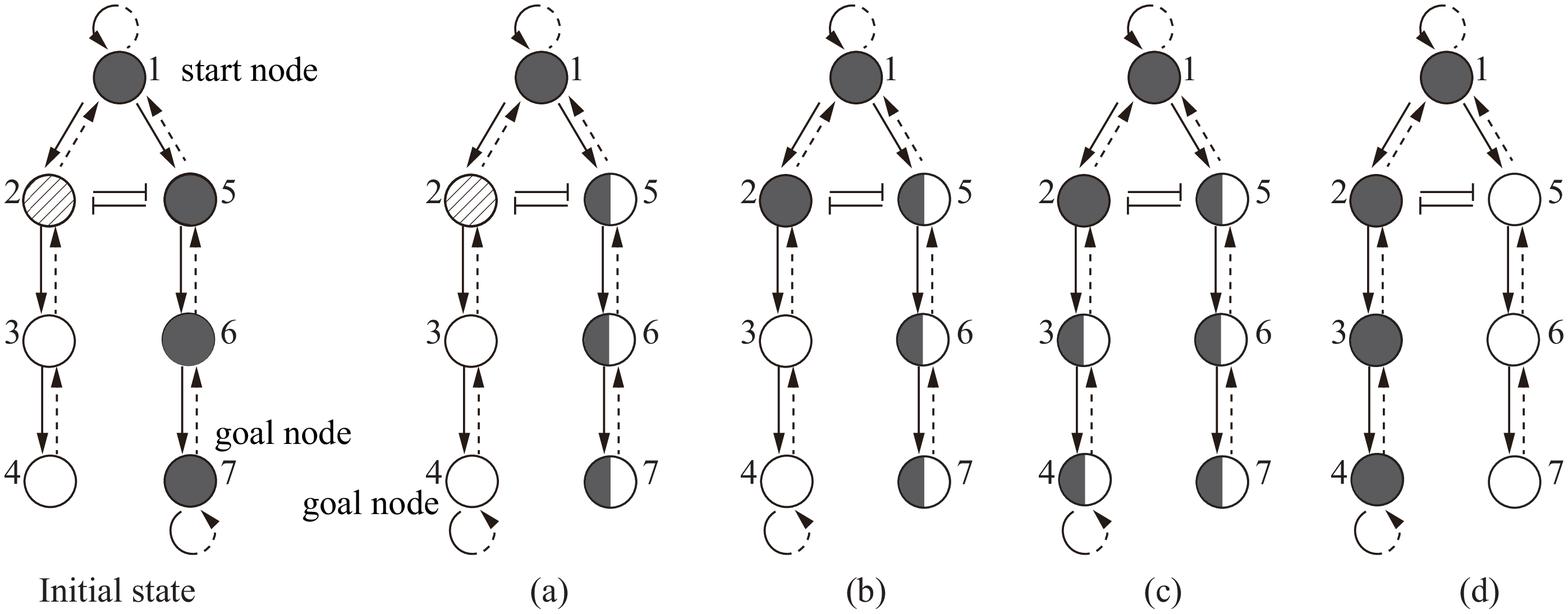}
\end{center}
 \caption{Recovery process when the goal vertex is changed from $7$ to $4$.
 }
\label{fig:recover}
\end{figure}

The self-recovery process is explained by using a network shown in 
Fig.\ref{fig:recover}. 
For example, if the goal vertex is changed from $7$ to $4$
[Fig.\ref{fig:recover}(a)], the 
state of nodes $6^-$, $5^-$, and 
$4^-$ successively becomes INC due to (C1),
and the state of node $2^\pm$ is changed from SSO to LSO due to (C3)
[Fig.\ref{fig:recover}(b)].  
Due to (C1), the state of nodes $3^+$, $4^+$, $4^-$, and $3^-$
successively becomes LSO, and the desired
path is found [Fig.\ref{fig:recover}(c)(d)]. It should be noted that the
LSO state of node $2^-$ needs to be  
maintained during the process (c) to (d). 
This means that the transient time from LSO to INC
should be chosen to be large enough compared to the LSO wave speed and network
size by taking $\mu_1$ near the limit point in Fig.\ref{fig:bif}. 

If the graph contains a cyclic loop that does not contain the start vertex or the goal
vertex, the system may fail to find the connecting path between the start and
goal nodes, since an LSO loop along the cyclic loop is stable, and the
system may choose it. Thus, we assumed (S6).

\acknowledgements
This work was supported by 
a Grant-in-Aid for Scientific Research on
Innovative Areas ``Neural creativity for communication'' (No.\ 4103) 
(21120003 and 21120005) of MEXT, Japan and 
a Grant-in-Aid for Scientific Research (C) (25400199) of JSPS, Japan.

\end{document}